\documentclass[a4,11pt,oneside]{article}

\usepackage[margin=3cm]{geometry}
\usepackage[english]{babel}
\usepackage[latin1]{inputenc}
\usepackage{indentfirst}
\usepackage{amsmath, amssymb, mathrsfs, dsfont, mathtools, empheq}
\usepackage{amsthm}
\usepackage{graphicx}
\usepackage{comment}

\newcommand{\eq}{\begin{equation}}
\newcommand{\eeq}{\end{equation}}

\newcommand{\R}{\mathbb R}
\newcommand{\N}{\mathbb N}
\newcommand{\E}{\mathbb E}
\newcommand{\Prob}{\mathbb P}
\newcommand{\1}{\mathds 1}

\newcommand{\dd}{\mathrm d}
\newcommand{\eps}{\varepsilon}
\newcommand{\bigchi}{\mathcal X}

\newcommand{\DD}{\mathscr D}
\newcommand{\xv}{\boldsymbol{x}}
\newcommand{\xiv}{\boldsymbol{\xi}}
\newcommand{\Exv}{\E_{\xv}}
\newcommand{\Probxv}{\Prob_{\xv}}
\newcommand{\Ex}{\E_x}
\newcommand{\Probx}{\Prob_x}
\newcommand{\Exiv}{\E_{\xiv}}
\newcommand{\Exi}{\E_\xi}
\newcommand{\x}{\boldsymbol{x}}
\newcommand{\w}{\boldsymbol{w}}

\theoremstyle{plain}
\newtheorem{thm}{Theorem} [section]
\newtheorem{lem}[thm]{Lemma}
\newtheorem{prop}[thm]{Proposition}
\newtheorem{cor}[thm]{Corollary}

\theoremstyle{definition}
\newtheorem{df}[thm]{Definition}
\newtheorem{rk}[thm]{Remark}

\title{A mean-field monomer-dimer model with randomness.\\
Exact solution and rigorous results.}
\author{Diego Alberici, Pierluigi Contucci, Emanuele Mingione}
\date{\today}

\begin{document}
\maketitle

\begin{abstract}
Independent random monomer activities are considered on a mean-field monomer-dimer model.
Under very general conditions on the randomness the model is shown to have
a self-averaging pressure density that obeys an exactly solvable variational principle. The
dimer density is exactly computed in the thermodynamic limit and shown to be
a smooth function.
\end{abstract}

\vspace{8pt}


\section{Introduction}

In this paper we study a mean-field monomer-dimer model with randomness in the monomer activities.
The model describes, in the mean-field approximation, the equilibrium properties of a system of diatomic molecules
(see e.g. \cite{Ro,FR,C,HL}) depositing on an inhomogeneous lattice. The inhomogeneity is modelled by introducing
a quenched randomness on the site activity, in the same spirit that the random field Ising model \cite{belanger, SDP, bovier}
describes inhomogeneity for ferromagnets.

The main feature of any monomer-dimer system is the hard-core interaction among the diatomic particles:
different atoms cannot deposit on the same site, due to the repulsivity of the van der Waals potential at short distances.
By adding a random monomer activity, the present work is intended to be a first step toward a more realistic description
of the physical phenomenon of adsorption which includes randomness at many levels and in particular
also for the interactions among particles.
Beside the short distance repulsion modelled by the hard-core constraint, the physical system
displays also an attractive interaction among particles when they are beyond the equilibrium distance
of the van der Waals potential \cite{FR,C,HLliq}. In \cite{ACM,ACMepl} an attraction coupling was introduced and
the model exactly solved in the mean-field lattice, dysplaying a first order phase transition.
Clearly a fully realistic model should include quenched random interactions as well, in the same
spirit that leads to the spin glass description of magnetic systems \cite{EA}.
From this point of view, the model studied here plays the role of the zero-coupling random field spin model
for the mean-field spin glass \cite{SK, MPV} with the essential difference that it is still an interacting model due to the
presence of the hard-core repulsion and, by consequence, the Boltzmann-Gibbs measure does not factorise. A rigorous
solution of the full model relies on the exact solvability of its zero coupling limit \cite{G}.

Concerning the mean-field nature of the model investigated here, it is expected to be an
approximation for models in the finite-dimensional lattices, in the same sense that occurs for
the ferromagnetic spin systems and their quenched versions \cite{tom, franz}.

From the mathematical point of view the model presented here admits also the interpretation of a pair matching
problem with random weighted sites (see e.g. \cite{Gi} for an overview of matching problems).
For a different way of introducing randomness in monomer-dimer systems see \cite{AC}, where a model on locally tree-like random graphs is solved. The combinatorial problem of perfect matchings on random graphs, already solved in \cite{KS, BLS}, corresponds the zero-temperature limit of the latter monomer-dimer model.

In the present work, where the only interaction considered is of hard-core type, our approach builds on the fundamental results by Heilmann and Lieb \cite{HL,HLprl} and their general proof of the absence of phase transitions.
Our main result is the exact solution of the model with \textit{i.i.d.}$\!$ randomness on the monomer activities $x_i$'s.
Precisely we prove that, in the thermodynamic limit, the pressure density exists under very general conditions on the probability distribution and it is given by a variational principle of elementary nature, i.e. the maximisation of a function $\Phi$ on the positive real line, where
\[ \Phi(\xi) \;=\; -\frac{\xi^2}{2w} \,+\, \E_x[\log(\xi+x)] \;,\quad \xi\geq0 \;.\]
The pressure density turns out to be a smooth function of the dimer activity $w$. The dimer density $d=d(w)$ is therefore a smooth function too and it is described by the formula
\[ d \; = \; \frac{(\xi^*)^2}{2w} \]
where $\xi^*$ is the unique positive solution of the fixed point equation
\[ \xi^* \, = \; w\; \Ex\bigg[\frac{1}{\xi^*+x}\bigg] \;.\]

The problem, otherwise expected to be difficult due the hard-core interaction among dimers,
becomes accessible with the use of a Gaussian representation for the partition function.
As we will show in the next section, this representation ``factorizes'' the hard-core constraints in the same way as the Hubbard-Stratonovich transform decouples other types of two-body interactions. In our case this simplification comes with a technical difficulty: the Gaussian representation for the partition function displays an integrand with negative values and possible singular contributions. However, a careful application of the uniform law of large numbers and the Laplace method allows us to overcome the reach a fully rigorous control and obtain the exact solution.

It is interesting to emphasize that the Gaussian representation for the partition function is able to capture the essence of the
Heilmann-Lieb recursion relation that is the main tool to solve many monomer-dimer models \cite{HL,AC}. We
show in fact that this recursion relation reduces to integration by parts of the Gaussian measure. In the present paper
the Heilmann-Lieb recursion relation and some technical methods for martingales (like the Azuma's inequality) are the ingredients used to prove the self-averaging of the pressure density.

The paper is organised as follows.
In the section \ref{sec: gauss} we describe the Gaussian representation for the partition function of a general monomer-dimer model and we deduce the Heilmann-Lieb recursion.
In the section \ref{sec: random monomer} we solve the monomer-dimer model on the complete graph with \textit{i.i.d.}$\!$ random monomer activities; in particular we compute the pressure density in the theorem \ref{thm: main} and the dimer density in the corollary \ref{cor: dimer density}.
In the section \ref{sec: self-av} we show, under suitable assumptions, that the free energy density of a monomer-dimer model with independent random activities is self-averaging. The appendix collects the main technical results used in this paper, in order to make it self contained.

\section{Gaussian representation for monomer-dimer models} \label{sec: gauss}
In this section we recall the definition of a monomer-dimer model with pure hard-core interaction and we show how to write its partition function as a Gaussian expectation. This representation, which will be extensively used in this work, was
first proposed in \cite{V} and is an immediate consequence of the
Wick-Isserlis formula for Gaussian moments. As a first application we show in this section that the well-known Heilmann-Lieb recursion formula \cite{HL} for monomer-dimer models corresponds in fact to a Gaussian integration by parts.

\begin{df}
Let $G=(V,E)$ be a finite simple graph.
A \textit{dimer configuration} (or \textit{matching}) on $G$ is a set $D$ of pairwise non-incident edges (called \textit{dimers}).
The associated set of dimer-free vertices (called \textit{monomers}) is denoted by $M_G(D)$.
In other terms a dimer configuration $D$ on $G$ is a partition of a certain set $A\subseteq V$ into pairs belonging to $E\,$:
\eq \begin{split}
& D = \big\{ \{i_1,i_2\}, \dots, \{i_{|A|-1},i_{|A|}\} \big\} \\[2pt]
& \text{with }\ \{i_1,i_2,\dots,i_{|A|}\}=A\ \text{ and }\ \{i_s,i_{s+1}\}\in E \;;
\end{split} \eeq
and the associated monomer set is $M_G(D)=V\smallsetminus A\,$.\\[2pt]
Denote by $\DD_G$ the space of all possible dimer configurations on the graph $G\,$.
A \textit{monomer-dimer model (with pure hard-core interaction)} on $G$ is obtained by assigning a monomer weight $x_i>0$ to each vertex $i\in V$, a dimer weight $w_{ij}\geq0$ to each edge $ij\equiv\{i,j\}\in E$ and introducing the following Gibbs probability measure on $\DD_G\,$:
\eq
\mu_{G}(D) \,:=\, \frac{1}{Z_G}\; \prod_{ij\in D}w_{ij} \prod_{\,i\in M_G(D)}\!\!\!x_i\, \quad\forall\,D\in\DD_G \;,
\eeq
where $Z_G := \sum_{D\in\DD_G} \prod_{ij\in D}w_{ij} \prod_{i\in M_G(D)}x_i$ is the normalizing factor, called \textit{partition function}.
\end{df}

The following remark shows that, when the weights are kept so general, it is sufficient (and convenient) to work on a complete graph.

\begin{rk} \label{rk: complete}
Consider the complete graph $K_N$, with vertex set $\{1,\dots,N\}$ and edge set made of all possible pairs of vertices.
Because of the lack of geometric structure the space of dimer configurations $\DD_N\equiv\DD_{K_N}$ simplifies; precisely $D\in\DD_N$ if and only if
\eq
D = \big\{ \{i_1,i_2\}, \dots, \{i_{|A|-1},i_{|A|}\} \big\}\ \text{ with }\ \{i_1,i_2,\dots,i_{|A|}\}=A
\eeq
for a certain set of vertices $A\subseteq\{1,\dots,N\}$, and the monomer set associated to $D$ is $M_N(D) \equiv M_{K_N}(D) = \{1,\dots,N\}\smallsetminus A$.\\
On the other hand any monomer-dimer model on a graph $G=(V,E)$ with $N$ vertices can be thought as a monomer-dimer model on the complete graph $K_N$. Indeed the measure $\mu_G$ is equivalent to a measure $\mu_N\equiv\mu_{K_N}$ by setting $w_{ij}:=0$ for all pairs $ij\notin E\,$. Precisely introducing these zero dimer weights it holds $Z_N\equiv Z_{K_N}=Z_G$ and
\[ \mu_N(D) = \begin{cases}
\mu_G(D) & \text{if }D\in\DD_G \\
0 & \text{if }D\in\DD_N\smallsetminus\DD_G
\end{cases} \;.\]
\end{rk}

The next proposition describes the Gaussian representation for the monomer-dimer model. Without loss of generality we work with the partition function $Z_N$ on the complete graph.

\begin{prop}[Gaussian representation] \label{prop: gauss repr}
The partition function of any monomer-dimer model over $N$ vertices can be written as
\eq \label{eq: gauss repr}
Z_N \,=\, \Exiv\bigg[\prod_{i=1}^N(\xi_i+x_i)\bigg] \;,
\eeq
where $\boldsymbol{\xi}=(\xi_1,\dots,\xi_N)$ is a Gaussian random vector with mean $0$ and covariance matrix $W=(w_{ij})_{i,j=1,\dots,N}$.
Here the diagonal entries $w_{ii}$ are arbitrary numbers, chosen in such a way that $W$ is a positive semi-definite matrix.
\end{prop}

\proof
As already noticed the dimer configurations on the complete graph are the partitions into pairs of all possible $A\subseteq\{1,\dots,N\}$, hence
\eq \label{eq: gauss repr proof1}
Z_N \,= \sum_{D\in\DD_N}\, \prod_{ij\in D}w_{ij} \prod_{\,i\in M_N(D)}\!\!\!\!x_i \,=
\sum_{A\subseteq\{1,\dots,N\}} \sum_{P\text{ partition}\atop\text{ of }A\text{ into pairs}}\, \prod_{ij\in P}w_{ij}\, \prod_{i\in A^c}x_i \;.
\eeq
Now choose $w_{ii}$ for $i=1,\dots,N$ such that the matrix $W=(w_{ij})_{i,j=1,\dots,N}$ is positive semi-definite\footnote{For example one can choose $w_{ii}\geq\sum_{j\neq i}w_{ij}$ for every $i=1,\dots,N$. $W$ can be diagonalized and has non-negative eigenvalues by the Gershgorin circle theorem, hence it is positive semi-definite.}.
Then there exists an (eventually degenerate) Gaussian vector $\boldsymbol{\xi}=(\xi_1,\dots,\xi_N)$ with mean $0$ and covariance matrix $W$. And by the Wick-Isserlis theorem (identity \eqref{eq: wick} in the theorem \ref{thm: gauss calculus})
\eq \label{eq: gauss repr proof2}
\Exiv\bigg[\prod_{i\in A}\xi_i\bigg] \;= \sum_{P\text{ partition}\atop\text{ of }A\text{ into pairs}} \prod_{ij\in P}w_{ij} \;.
\eeq
Substituting \eqref{eq: gauss repr proof2} into \eqref{eq: gauss repr proof1} one obtains
\eq
Z_N \,=\, \Exiv\bigg[ \sum_{A\subseteq\{1,\dots,N\}}\, \prod_{i\in A}\xi_i\, \prod_{i\in A^c}x_i \bigg] \,=\,
\Exiv\bigg[ \prod_{i=1}^N(\xi_i+x_i) \bigg] \;.
\eeq
\endproof

\begin{rk} \label{rk: Hubb-Strat}
In some sense, the Gaussian representation \eqref{eq: gauss repr} ``factorises'' the hard-core constraints among dimers in the same way as the Hubbard-Stratonovich transform decouples the two-body interactions in the Ising model.
For the sake of clarity, consider a generic Ising partition function:
\[  Z_N^\textup{Ising} \,=\,
\sum_{\,\sigma\in\{\pm1\}^N} e^{\sum_{1\leq i<j\leq N} J_{ij}\sigma_i\sigma_j}\ e^{\sum_{i=1}^N h_i\sigma_i}
\; \]
Set $J_{ij}=J_{ji}$ and $J_{ii}\geq\sum_{j\neq i}|J_{ij}|$ so that $J:=(J_{ij})_{i,j=1,\dots,N}$ is a real positive semi-definite matrix, by the Gershgorin's circle theorem.
The Hubbart-Stratonovich transform simply relies on the computation of the Gaussian moment generating function:
\[ e^{\frac{1}{2}\,\sum_{1\leq i,j\leq N} J_{ij}\sigma_i\sigma_j} \,=\, \E_{\boldsymbol\xi'}\big[e^{\sum_{i=1}^N\xi_i\sigma_i}\big] \]
where $\boldsymbol\xi'=(\xi'_1,\dots,\xi'_N)$ is a Gaussian random vector with mean $0$ and covariance matrix $J$.
Therefore the problem factorises and one obtains
\[ Z_N^\textup{Ising} \,=\,
e^{-\frac{1}{2}\sum_{i=1}^N\!J_{ii}}\ \E_{\boldsymbol\xi'}\bigg[ \prod_{i=1}^N 2 \cosh(\xi'_i+h_i) \bigg] \,.\]
\end{rk}

As an application of the Gaussian representation we show that the well-know Heilmann-Lieb recursion \cite{HL} for the partition function of monomer-dimer models can be proved by means of a Gaussian integration by parts.

\begin{prop}[Heilmann-Lieb recursion] \label{prop: HL rec}
Let $G=(V,E)$ be a finite simple graph and consider a monomer-dimer model on $G$. Fix $i\in V$ and look at its adjacent vertices $j\sim i$, then it holds
\eq \label{eq: HL rec}
Z_G \,=\, x_i\,Z_{G-i} \,+\, \sum_{j\sim i}\,w_{ij}\,Z_{G-i-j} \;.
\eeq
Here $G-i$ denotes the graph obtained from $G$ deleting the vertex $i$ and all its incident edges.
\end{prop}

\proof[Proof using Gaussian integration by parts]
Set $N:=|V|$. Introduce zero dimer weights $w_{hk}=0$ for the pairs $hk\notin E$, so that $Z_G=Z_N$ (see remark \ref{rk: complete}).
Following proposition \ref{prop: gauss repr}, introduce an $N$-dimensional Gaussian vector $\boldsymbol{\xi}$ with mean $0$ and covariance matrix $W$. Then write the identity \eqref{eq: gauss repr} isolating the vertex $i\,$:
\eq \label{eq: HL rec proof1}
Z_G \,=\, \Exiv\bigg[\prod_{k=1}^N(\xi_k+x_k)\bigg] \,=\,
x_i\;\Exiv\bigg[\prod_{k\neq i}(\xi_k+x_k)\bigg] \,+\, \Exiv\bigg[\xi_i\,\prod_{k\neq i}(\xi_k+x_k)\bigg] \;.
\eeq
Now apply the Gaussian integration by parts (identity \eqref{eq: integr by parts} in the theorem \ref{thm: gauss calculus}) to the second term on the r.h.s. of \eqref{eq: HL rec proof1}:
\eq \label{eq: HL rec proof2}
\Exiv\bigg[\xi_i\,\prod_{k\neq i}(\xi_k+x_k)\bigg] \,=\,
\sum_{j=1}^N\, \Exiv[\xi_i\xi_j]\; \Exiv\bigg[ \frac{\partial}{\partial \xi_j}\prod_{k\neq i}(\xi_k+x_k) \bigg] \,=\,
\sum_{j\neq i}\, w_{ij}\; \Exiv\bigg[\prod_{k\neq i,j}(\xi_k+x_k) \bigg] \;.
\eeq
Notice that summing over $j\neq i$ in the r.h.s. of \eqref{eq: HL rec proof2} is equivalent to sum over $j\sim i$, since by definition $w_{ij}=0$ if $ij\notin E$. Substitute \eqref{eq: HL rec proof2} in \eqref{eq: HL rec proof1}:
\eq \label{eq: HL rec proof3}
Z_G \,=\,
x_i\;\Exiv\bigg[\prod_{k\neq i}(\xi_k+x_k)\bigg] \,+\, \sum_{j\sim i}\, w_{ij}\; \Exiv\bigg[\prod_{k\neq i,j}(\xi_k+x_k) \bigg] \;.
\eeq
To conclude observe that $(\xi_k)_{k\neq i}$ is an $(N-1)$-dimensional Gaussian vector with mean $0$ and covariance $(w_{hk})_{h,k\neq i}$. Hence by proposition \ref{prop: gauss repr}
\eq \label{eq: HL rec proof4}
Z_{G-i} \,=\, \Exiv\bigg[\prod_{k\neq i}(\xi_k+x_k)\bigg] \;.
\eeq
And similarly
\eq \label{eq: HL rec proof5}
Z_{G-i-j} \,=\, \Exiv\bigg[\prod_{k\neq i,j}(\xi_k+x_k)\bigg] \;.
\eeq
Substitute the identities \eqref{eq: HL rec proof4}, \eqref{eq: HL rec proof5} into \eqref{eq: HL rec proof3} to obtain the identity \eqref{eq: HL rec}.
\endproof

\section{Monomer-dimer model with random monomer weights} \label{sec: random monomer}
In this section we fix a uniform dimer weight on the complete graph, while we choose i.i.d. random monomer weights. Under quite general integrability hypothesis, we show that this model is exactly solvable and it does not present a phase transition (in agreement with the general results by Heilmann and Lieb \cite{HL, HLprl}).

Let $w>0$. Let $x_i>0$, $i\in\N$, be \textit{independent identically distributed} random variables. In order to keep the logarithm of the partition function of order $N$, a normalization of the dimer weight as $w/N$ is needed. Therefore during all this section we will denote
\eq \label{eq: Z unif w}
Z_N \,=\, \sum_{D\in\DD_N} \big(\,\frac{w}{N}\,\big)^{|D|} \!\!\prod_{\,i\in M_N(D)}\!\!\!\!x_i \;.
\eeq
$\mu_N$ will denote the corresponding Gibbs measure and $\langle\,\cdot\,\rangle_N$ will be the expected value with respect to $\mu_N$.
Notice that now the partition function is a random variable and the Gibbs measure is a random measure.

\begin{rk} \label{rk: gauss repr unif w}
Since the dimer weight is uniform, the Gaussian representation of \eqref{eq: Z unif w} simplifies:
\eq \label{eq: gauss repr unif w}
Z_N \,=\, \Exi\bigg[\prod_{i=1}^N(\xi+x_i)\bigg] \;,
\eeq
where $\xi$ is a $1$-dimensional Gaussian random variable with mean $0$ and variance $w/N$.\\[1pt]
Indeed by proposition \ref{prop: gauss repr}, $Z_N = \Exiv\big[\prod_{i=1}^N(\xi_i+x_i)\big]$ where $\boldsymbol{\xi}=(\xi_1,\dots,\xi_N)$ is an $N$-dimensional Gaussian random vector with mean $0$ and constant covariance matrix\footnote{It is important to notice that setting also the diagonal entries to $w/N$, the resulting matrix is positive semi-definite: $\sum_{i=1}^N\sum_{j=1}^N(w/N)\,\alpha_i\alpha_j = (w/N)\,\big(\sum_{i=1}^N\alpha_i\big)^2 \geq 0\,$ for every $\alpha\in\R^N$.
} $(w/N)_{i,j=1,\dots,N}\,$.
It is easy to check that $\boldsymbol{\xi}$ has the same joint distribution of the constant random vector $(\xi,\dots,\xi)$.
Therefore the identity \eqref{eq: gauss repr unif w} follows.
\end{rk}

\begin{rk}
Keeping in mind the remark \ref{rk: Hubb-Strat}, one can observe the analogy among the formula \eqref{eq: gauss repr unif w} and the partition function of the Curie-Weiss random field model (see e.g. \cite{SP,A,SW}), that is
\eq \label{eq: gauss CW}
 Z_N^\textup{Curie-Weiss} \,=\, e^{\frac{J}{2}}\ \E_{\xi'}\bigg[\prod_{i=1}^N 2\cosh(\xi'+h_i) \bigg]
\eeq
where $\xi'$ is a $1$-dimensional Gaussian random variable with mean $0$ and variance $J/N\in\R_+$.\\
By the way, we want to stress the fact that the Laplace method applies directly to formula \eqref{eq: gauss CW}, while the presence of negative and singular contributions in \eqref{eq: gauss repr unif w} will require a supplementary work in order to study the asymptotic behaviour.
\end{rk}

Let us rewrite \eqref{eq: gauss repr unif w} as an explicit integral in $\dd\xi$:
\eq  \label{eq: gauss repr unif w 1}
Z_N \,=\, \frac{\sqrt{N}}{\sqrt{2\pi w}}\, \int_\R e^{-\frac{N}{2w}\,\xi^2}\, \prod_{i=1}^N(\xi+x_i) \;\dd\xi  \;.
\eeq

\begin{thm} \label{thm: main}
Let $w>0$. Let $x_i>0,\,i\in\N$ be i.i.d. random variables. Denote by $x$ a random variable distributed like $x_i$; suppose that $\Ex[x]<\infty$ and $\Ex[(\log x)^2]<\infty$. Then:
\eq \label{eq: var princ}
\exists\ \lim_{N\to\infty} \frac{1}{N}\,\Exv[\,\log Z_N] \;=\; \sup_{\xi\geq0}\Phi(\xi)
\eeq
where
\eq \label{eq: Phi}
\Phi(\xi) \,:=\, -\frac{\xi^2}{2w} + \Ex[\,\log(\xi+x)] \quad\forall\,\xi\geq0 \;.
\eeq
Furthermore the function $\Phi$ attains its maximum at a unique point $\xi^*$. $\xi^*$ is the only solution in $\,[0,\infty[\,$ of the fixed point equation
\eq \label{eq: fixed point}
\xi^* \,=\, \Ex\bigg[\frac{w}{\xi^*+x}\bigg] \;.
\eeq
Thus the following bounds hold:
\eq \label{eq: bounds xi0}
\frac{-\Ex[x]+\sqrt{\Ex[x]^2+4w}}{2}\,\lor\,\sup_{t>0}\frac{-t+\sqrt{t^2+4w\,\Probx(x\leq t)}}{2} \;\leq\; \xi^* \;\leq\;
\sqrt{w}\,\land\,\Ex\bigg[\frac{w}{x}\bigg] \;.
\eeq
\end{thm}

In consequence of the theorem \ref{thm: main} it is not hard to prove that the system does not present a phase transition in the parameter $w>0$.
It is also easy to compute the main macroscopic quantity of physical interest, that is the \textit{dimer density}, in terms of the positive solution $\xi^*$ of the fixed point equation \eqref{eq: fixed point}.
Therefore we state the following two corollaries before starting to prove the theorem.

\begin{cor} \label{cor: smooth}
In the hypothesis of the theorem \ref{thm: main}, consider the limiting pressure density function $p(w):=\lim_{N\to\infty}\frac{1}{N}\,\Exv\big[\log Z_N(w)\big]$ for all $w>0\,$. Then $p\in C^\infty\big(\,]0,\infty[\,\big)\,$.
\end{cor}

\proof
By the theorem \ref{thm: main} $p(w)=\Phi(w,\xi^*)$, where $\Phi(w,\xi)=-\xi^2/(2w)+\Ex[\,\log(\xi+x)]$ and $\xi^*=\xi^*(w)$ is \textit{the only} positive solution of the equation $F(w,\xi)=0$ with $F:=\frac{\partial\Phi}{\partial \xi}\,$.\\
$F$ is a smooth function on $]0,\infty[\,\times\,]0,\infty[\,$, because $\Phi$ is smooth as it will be proven in the lemma \ref{lem: Phi}. In addition $\frac{\partial F}{\partial \xi}(w,\xi^*)\neq0$ for all $w>0$, by the lemma \ref{lem: Phi} equation \eqref{eq: Phi''}.\\
As a consequence, by the implicit function theorem (see e.g. \cite{Ru}), $\xi^*$ is a smooth function of $w\in\,]0,\infty[\,$.
Hence, by composition, also $p(w)=\Phi\big(w,\xi^*(w)\big)$ is a smooth function of $w\in\,]0,\infty[\,$.
\endproof

\begin{cor} \label{cor: dimer density}
In the hypothesis of the theorem \ref{thm: main}, the limiting dimer density
\[ d:=\lim_{N\to\infty}\frac{1}{N}\,\Exv\big[\big\langle\,|D|\,\big\rangle_{\!N\,}\big] \]
can be computed as
\eq \label{eq: dimer density}
d \,=\, w\,\frac{\dd\, p}{\dd w} \,=\, \frac{(\xi^*)^2}{2w} \;.
\eeq
\end{cor}

\proof
Set $p_N:=\frac{1}{N}\log Z_N$ and perform the change of parameter $w=:e^h$. Clearly $\frac{\dd}{\dd h} = w\,\frac{\dd}{\dd w}$ and it is easy to check that
\[ \frac{\dd\, \Exv[p_N]}{\dd h} \,=\, \Exv\big[\big\langle\,|D|\,\big\rangle_{\!N\,}\big] \;.\]
By the theorem \ref{thm: main} and its corollary \ref{cor: smooth}, $\Exv[p_N]$ converges pointwise to a smooth function $p$ as $N\to\infty$ for all values of $h\in\R$. A standard computation shows that $\Exv[p_N]$ is a convex function of $h$. Therefore
\[ \frac{\dd\, \Exv[p_N]}{\dd h} \,\xrightarrow[N\to\infty]\, \frac{\dd\, p}{\dd h} \;.\]
Since $p(h)=\Phi\big(h,\xi^*(h)\big)$, where $\xi^*$ is the critical point of $\Phi$ and is a smooth function of $h$, it is easy to compute
\[ \frac{\dd\, p}{\dd h}(h) \,=\,\frac{\partial\Phi}{\partial h}(h,\xi^*) \,+\, \underbrace{\frac{\partial\Phi}{\partial \xi}(h,\xi^*)}_{=\,0}\,\frac{\dd \xi^*}{\dd h}(h) \,=\, \frac{(\xi^*)^2}{2\,e^h} \;. \qedhere\]
\endproof

Now let us start to prove the theorem \ref{thm: main}. The logic structure of the proof is divided in three main parts.
First we study the basic properties of the function $\Phi$. Then we use the uniform law of large numbers and other observations to show that for large $N$ the integrated function in \eqref{eq: gauss repr unif w 1} can be well approximated by $e^{N\Phi}$. Finally we will be able to exploit the Laplace's method in order to compute a lower and an upper bound for $\frac{1}{N}\,\Exv[\,\log Z_N]\,$.

\begin{lem} \label{lem: Phi}
$\Phi$ is continuous on $[0,\infty[\,$, it is smooth on $]0,\infty[\,$ and the derivatives can be taken inside the expectation. In particular for all $\xi>0$ it holds
\begin{gather}
\label{eq: Phi'}
\Phi'(\xi) \,=\,  -\frac{\xi}{w} + \Ex\bigg[\frac{1}{\xi+x}\bigg] \;;\\[2pt]
\label{eq: Phi''}
\Phi''(\xi) \,=\, -\frac{1}{w} - \Ex\bigg[\frac{1}{(\xi+x)^2}\bigg] < 0 \;.
\end{gather}
As a consequence $\Phi$ has exactly one critical point $\xi^*$ in $]0,\infty[\,$,
that is the equation \eqref{eq: fixed point} has exactly one solution in $\,]0,\infty[\,$.
$\xi^*$ is \textup{the only global maximum point} of $\Phi$ on $[0,\infty[\,$. 
\end{lem}

\proof
\textbf{I.} First of all $\Phi(\xi)$ is well-defined for all $\xi\geq0$. Indeed for $\xi>0$
\[ \log(\xi+x) \,\begin{cases}
\,\leq\, \xi+x-1 \,\in L^1(\Probx) \\
\,\geq\, 1-\frac{1}{\xi+x} \,\geq\, 1-\frac{1}{\xi} \,\in L^1(\Probx)
\end{cases} \;;\]
while for $\xi=0$, $\Ex[|\!\log x|]\leq\Ex[(\log x)^2]^{1/2}<\infty$ by the H\"older inequality.\\
$\Phi$ is continuous at $\xi=0$ by monotone convergence: $\log(\xi+x)$ decreases to $\log x$ as $\xi\searrow0$ and $\Ex[\,\log(\xi+x)]<\infty\,$.\\
Let now $\xi>0$ and let $\delta>0$ such that $\xi-\delta>0$.
The first derivative of $\Phi$ at $\xi$ can be computed inside the expectation, obtaining \eqref{eq: Phi'}, since the difference quotient of $\xi\mapsto\log(\xi+x)$ satisfies the dominated convergence hypothesis. Indeed for all $\xi'\in\,]\xi-\delta,\xi+\delta[$
\[ \bigg|\frac{\log(\xi'+x)-\log(\xi+x)}{\xi'-\xi}\bigg| \,\leq\,
\sup_{\widetilde\xi\in[\xi,\xi']}\frac{1}{\widetilde\xi+x} \,\leq\,
\sup_{\widetilde\xi\in[\xi,\xi']}\frac{1}{\widetilde\xi} \;\leq\, \frac{1}{\xi-\delta} \ \in L^1(\Probx) \;.\]
Now the second derivative of $\Phi$ at $\xi$ can be computed inside the expectation, obtaining \eqref{eq: Phi''}, since the difference quotient of $\xi\mapsto\frac{1}{\xi+x}$ satisfies the dominated convergence hypothesis. Indeed for all $\xi'\in\,]\xi-\delta,\xi+\delta[$
\[ \bigg|\frac{\frac{1}{\xi'+x}-\frac{1}{\xi+x}}{\xi'-\xi}\bigg| \,\leq\,
\sup_{\widetilde\xi\in[\xi,\xi']}\frac{1}{(\widetilde\xi+x)^2} \,\leq\,
\sup_{\widetilde\xi\in[\xi,\xi']}\frac{1}{\big(\widetilde\xi\,\big)^2} \;\leq\, \frac{1}{(\xi-\delta)^2} \ \in L^1(\Probx) \;.\]
This reasoning can be iterated up to the derivative of any order, since $1/\big(\widetilde\xi+x\big)^k \leq 1/\big(\widetilde\xi\,\big)^k \leq 1/(\xi-\delta)^k \in L^1(\Probx)$ for all $\widetilde\xi\in\,]\xi-\delta,\xi+\delta[\,$ and all $k\geq1\,$. \\[2pt]
\textbf{II.} In virtue of \eqref{eq: Phi''} $\Phi$ is a strictly convex function on $\,]0,\infty[\,$. At the boundaries of this domain $\lim_{\xi\to0+}\Phi'(\xi)=\Ex[x^{-1}]>0$ and $\lim_{\xi\to\infty}\Phi'(\xi)=-\infty<0$ by \eqref{eq: Phi'} and monotone converge.
Therefore $\Phi$ has exactly one critical point $\xi^*$ in $\,]0,\infty[\,$ and 
it is the only global maximum point of $\Phi$.
\endproof

\begin{rk} \label{rk: bounds xi0}
Since $\xi^*$ satisfies the fixed point equation \eqref{eq: fixed point}, it is easy to obtain the bounds \eqref{eq: bounds xi0} for $\xi^*$.
Since $\xi^*>0$ and $x>0$,
\[ \xi^* = \Ex\bigg[\frac{w}{\xi^*+x}\bigg] \leq \frac{1}{\xi^*} \;\Rightarrow\; \xi^*\leq\sqrt{w} \;; \quad
\xi^* = \Ex\bigg[\frac{w}{\xi^*+x}\bigg] \leq \Ex\bigg[\frac{w}{x}\bigg] \;.\]
Using the Jensen inequality,
\[ \xi^* = \Ex\bigg[\frac{w}{\xi^*+x}\bigg] \geq \frac{w}{\xi^*+\Ex[x]} \;\Rightarrow\;
(\xi^*)^2+\xi^*\,\Ex[x]-w \geq 0 \;\Rightarrow\;
\xi^* \geq \frac{-\Ex[x]+\sqrt{\Ex[x]^2+4w}}{2} \;.\]
Finally, since $\xi^*+x>0$, it holds for all $t>0$
\[\begin{split}
\xi^* = \Ex\bigg[\frac{w}{\xi^*+x}\bigg] \geq \frac{w}{\xi^*+t}\,\Probx(x\leq t)
\;&\Rightarrow\; (\xi^*)^2+\xi^*\,t-w\,\Probx(x\leq t) \geq 0 \;\Rightarrow\\
&\Rightarrow\; \xi^* \geq \frac{-t+\sqrt{t^2+4w\,\Probx(x\leq t)}}{2} \;.
\end{split}\]
\end{rk}

\begin{lem} \label{lem: approx}
Define the random function
\eq \label{eq: PhiN}
\Phi_N(\xi) \,:=\, -\frac{\xi^2}{2w}+\frac{1}{N}\sum_{i=1}^N\,\log|\xi+x_i| \quad\forall\,\xi\in\R \;.
\eeq
This function is defined also for negative values of $\xi$ and it takes the value $-\infty$ at the random points $-x_1,\dots,-x_N$. It is important to observe that
\eq \label{eq: reflect PhiN}
\Phi_N(-\xi) \,<\, \Phi_N(\xi) \quad\forall\,\xi>0 \;.
\eeq
\textbf{i.} Let $0<M<\infty$. Then for all $\eps>0$
\eq \label{eq: approx [m,M]}
\Probxv\bigg(\, \forall\,\xi\!\in\![0,M]\;\ |\Phi_N(\xi)-\Phi(\xi)|<\varepsilon \,\bigg) \,\xrightarrow[N\to\infty]{}\,1 \;.
\eeq
\textbf{ii.} Let $0<m<M<\infty$. Then there exists $\lambda_{m,M}>0$ such that
\eq \label{eq: approx [-M,-m]}
\Probxv\bigg(\, \forall\,\xi\!\in\![m,M]\;\ \Phi_N(-\xi)<\Phi_N(\xi)-\lambda_{m,M} \,\bigg) \,\xrightarrow[N\to\infty]{}\,1 \;.
\eeq
\textbf{iii.} Let $C\in\R$. Then there exists $M_C>0$ such that
\eq \label{eq: approx ]-infty,-M] [M,infty[}
\Probxv\bigg(\, \forall\,\xi\!\in\![M_C,\infty[\,\;\ \Phi_N(\xi)<C\ \textup{ and }\ \Phi_N(\xi)<\varphi(\xi) \,\bigg) \,\xrightarrow[N\to\infty]{}\,1 \;;
\eeq
where $\varphi$ is the following deterministic function
\eq \label{eq: phi}
\varphi(\xi) \,:=\, -\frac{\xi^2}{2w} + \log\xi + \frac{1}{\xi}\,(\Ex[x]+1) \quad\forall\,\xi>0 \;.
\eeq
\end{lem}

Notice that $\Phi_N(\xi)-\Phi(\xi) = \frac{1}{N}\sum_{i=1}^N\log(\xi+x_i)-\Ex[\,\log(\xi+x)]$ for all $\xi>0$.
Since the $x_i,\,i\in\N$ are i.i.d., the basic idea behind the lemma \ref{lem: approx} is to approximate $\Phi_N$ with $\Phi\,$ by the law of large numbers. But this approximation is needed to hold at every $\xi$ at the same time, hence a \textit{uniform} law of large numbers is required.\\
To prove the theorem \ref{thm: main} it will be important to have found a good uniform approximation near the global maximum point $\xi^*$ of $\Phi$.
Far from $\xi^*$ instead such a uniform approximation cannot hold: for example $\Phi_N$ diverges to $-\infty$ at certain negative points, while, if the distribution of $x$ is absolutely continuous and satisfies some integrability hypothesis, it is possible to show that $\Phi(\xi)=-\frac{\xi^2}{2w}+\Ex[\,\log|\xi+x|]$ is continuous on $\R$. But fortunately, far from $\xi^*$, it will be sufficient for our purposes to bound suitably $\Phi_N$ from above.

\proof
\textbf{i.} For every $x>0$ the function $\xi\mapsto \log(\xi+x)$ is continuous on $[0,M]$ compact.
Moreover there is domination:
\[ \log(\xi+x) \,\begin{cases}\,\leq\, \log(M+x) \ \in L^1(\Probx) \\[2pt]
\,\geq\, \log x \ \in L^1(\Probx) \end{cases} \ \forall\,\xi\in[0,M] \;.\]
Therefore \eqref{eq: approx [m,M]} holds by the uniform weak law of large numbers (theorem \ref{thm: uwlln}). \\[2pt]
\textbf{ii.} Clearly $\log(\xi+x)>\log|-\xi+x|$ for all $\xi,x>0$. Furthermore an elementary computation shows that for all $\xi,x,\tau>0$
\[ \log(\xi+x)-\log|-\xi+x| \,\geq\, \tau \quad\Leftrightarrow\quad
\frac{e^\tau-1}{e^\tau+1}\,\xi \,\leq\, x \,\leq\, \frac{e^\tau+1}{e^\tau-1}\,\xi \;.\]
Therefore for all $\xi\in[m,M]$ and all $\tau>0$,
\eq \label{eq: approx proof2.1} \begin{split}
\Phi_N(\xi) - \Phi_N(-\xi) \,&=\, \frac{1}{N}\,\sum_{i=1}^N \big(\log(\xi+x_i)-\log|-\xi+x_i|\,\big) \,\geq\\
&\geq\, \frac{1}{N}\,\sum_{i=1}^N\, \tau\ \1\bigg(\frac{e^\tau-1}{e^\tau+1}\,\xi \,\leq\, x_i \,\leq\, \frac{e^\tau+1}{e^\tau-1}\,\xi\bigg) \,\geq\\
&\geq\, \tau\ \frac{1}{N}\,\sum_{i=1}^N\, \1\bigg(\frac{e^\tau-1}{e^\tau+1}\,M \,\leq\, x_i \,\leq\, \frac{e^\tau+1}{e^\tau-1}\,m\bigg) \;.
\end{split} \eeq
Set $I_{m,M}^\tau:=\big[\, \frac{e^\tau-1}{e^\tau+1}\,M \,,\, \frac{e^\tau+1}{e^\tau-1}\,m \,\big]$. Now by the weak law of large numbers, for all $\eps>0$
\eq \label{eq: approx proof2.2}
\Probxv\bigg(\, \frac{1}{N}\,\sum_{i=1}^N\, \1\big(x_i\in I_{m,M}^\tau\big) \,>\, \Probx\big(x\in I_{m,M}^\tau\big) - \eps \bigg) \,\xrightarrow[N\to\infty]{}\, 1 \;.
\eeq
Hence, using \eqref{eq: approx proof2.1} and \eqref{eq: approx proof2.2}, for all $\tau,\eps>0$
\eq \label{eq: approx proof2.3}
\Probxv\bigg(\, \Phi_N(\xi)-\Phi_N(-\xi) \,>\, \tau\, \big( \Probx(x\in I_{m,M}^\tau) - \eps \big) \bigg) \,\xrightarrow[N\to\infty]{}\, 1 \;.
\eeq
To conclude observe that $I_{m,M}^\tau\nearrow\;]0,\infty[\,$ (which is the support of the distribution of $x$) as $\tau\searrow0\,$. Hence there exists $\tau_0>0$ such that $\Probx(x\in I_{m,M}^{\tau_0})>0$. Choose $0<\eps_0<\Probx(x\in I_{m,M}^{\tau_0})$ and set
\[ \lambda_{m,M} \,:=\, \tau_0\, \big( \Probx(x\in I_{m,M}^{\tau_0}) - \eps_0 \big) \,> 0 \;.\]
Then \eqref{eq: approx [-M,-m]} follows from \eqref{eq: approx proof2.3}. \\[2pt]
\textbf{iii.} For all $\xi>0$ the following bound holds:
\eq \label{eq: approx proof3.1} \begin{split}
\Phi_N(\xi) \,&=\, -\frac{\xi^2}{2w} + \frac{1}{N}\sum_{i=1}^N\log(\xi+x_i) \,=\,
-\frac{\xi^2}{2w} + \log\xi + \frac{1}{N}\sum_{i=1}^N\log\big(1+\frac{x_i}{\xi}\big) \,\leq\\
&\leq\, -\frac{\xi^2}{2w} \,+\, \log\xi \,+\, \frac{1}{\xi}\;\frac{1}{N}\sum_{i=1}^N\,x_i \;.
\end{split} \eeq
Now by the weak law of large numbers (no uniformity in $\xi$ is needed here), for all $\eps>0$
\eq \label{eq: approx proof3.2}
\Probxv\bigg( \frac{1}{N}\sum_{i=1}^N\,x_i \,<\, \Ex[x] + \varepsilon \bigg) \,\xrightarrow[N\to\infty]{}\,1 \;.
\eeq
Hence, using \eqref{eq: approx proof3.1} and \eqref{eq: approx proof3.2}, for all $0<\eps<1$
\eq \label{eq: approx proof3.3}
\Probxv\bigg( \forall\,\xi\!>\!0\ \ \Phi_N(\xi) < \varphi(\xi) \bigg) \,\xrightarrow[N\to\infty]{}\,1 \;.
\eeq
Furthermore it holds $\varphi(\xi) \to -\infty$ as $\xi\to\infty\,$. Hence for all $C\in\R$ there exists $M_C>0$ such that
\eq \label{eq: approx proof3.4}
\varphi(\xi) < C \quad\forall\,\xi>M_C \;.
\eeq
In conclusion \eqref{eq: approx ]-infty,-M] [M,infty[} follows from \eqref{eq: approx proof3.3} and \eqref{eq: approx proof3.4}.
\endproof

\begin{lem} \label{lem: bound ZN}
There exists a constant $C_0<\infty$ such that
\eq \label{eq: bound ZN}
\Exv\bigg[\bigg(\frac{\log Z_N}{N}\bigg)^{\!\!2\,}\bigg] \,\leq\, C_0 \quad\forall\,N\in\N \;.
\eeq
\end{lem}

\proof
Since $x\mapsto(\log x)^2$ is concave for $x\geq e$, the Jensen inequality can be used as follows:
\eq \label{eq: bound ZN proof1.1} \begin{split}
\Exv\big[ (\log Z_N)^2\; \1(Z_N\geq e) \big] \,&=\,
\Exv\big[ (\log Z_N)^2 \,\big|\,Z_N\geq e\big]\ \Probxv(Z_N\geq e) \,\leq\\[2pt]
&\leq\, \big( \log\Exv\big[Z_N\,\big|\,Z_N\geq e\big] \,\big)^2\ \Probxv(Z_N\geq e) \,=\\[2pt]
&=\, \bigg( \log\frac{\Exv\big[Z_N\; \1(Z_N\geq e) \big]}{\Probxv(Z_N\geq e)} \bigg)^{\!2}\; \Probxv(Z_N\geq e) \,\leq\\[2pt]
&\leq\, 2\,\big( \log\Exv\big[Z_N\big] \big)^2 \,+\, 2 \max_{p\in[0,1]}(\log p)^2\,p \;.
\end{split} \eeq
Since the $x_i,\,i\in\N$ are i.i.d. $\Exv[Z_N]$ equals a deterministic partition function with uniform weights. Hence it is easy to bound it as follows:
\eq \label{eq: bound ZN proof1.2} \begin{split}
\Exv\big[Z_N\big] \,&=
\sum_{D\in\DD_N} \bigg(\frac{w}{N}\bigg)^{\!|D|}\, \Ex[x]^{|M(D)|} \,\leq\,
\sum_{d=0}^{|E_N|} {|E_N| \choose d}\, \bigg(\frac{w}{N}\bigg)^{\!d}\; \Ex[x]^{N-2d} \,=\\
&=\,\Ex[x]^N\, \bigg( 1+\frac{w}{N}\,\Ex[x]^{-2} \bigg)^{\!|E_N|} \,\leq\,
\Ex[x]^N\, \exp\bigg(\frac{N-1}{2}\;\frac{w}{\Ex[x]^2}\bigg)
\end{split} \eeq
(here $|E_N|=\frac{N(N-1)}{2}$ denotes the number of edges in the complete graph over $N$ vertices).
Therefore, substituting \eqref{eq: bound ZN proof1.2} into \eqref{eq: bound ZN proof1.1},
\eq \label{eq: bound ZN proof1.3}
\Exv\big[ (\log Z_N)^2\; \1(Z_N\geq e) \big] \,\leq\,
2\,N^2\,\bigg( \log\Ex[x] + \frac{w}{2\,\Ex[x]^2} \bigg)^{\!2} +\,  2 \max_{p\in[0,1]}(\log p)^2\,p \;.
\eeq
It remains to deal with the case $Z_N<e\,$. When $1<Z_N<e$, it holds $0<\log Z_N<1$ hence trivially
\eq \label{eq: bound ZN proof2.1}
\Exv\big[ (\log Z_N)^2\; \1(1<Z_N<e) \big] \,\leq\,
\Exv\big[ (\log e)^2\; \1(1<Z_N<e) \big] \,\leq\, 1 \;.
\eeq
When instead $Z_N\leq1$, it holds $\log Z_N\leq0$ hence we need a lower bound for $Z_N$. For example, considering only the configuration with no dimers, $Z_N\geq\prod_{i=1}^Nx_i\,$. Therefore:
\eq \label{eq: bound ZN proof2.2} \begin{split}
\Exv\big[ (\log Z_N)^2\; \1(Z_N\leq1) \big] \,&\leq\,
\Exv\bigg[ \bigg(\log\prod_{i=1}^Nx_i\bigg)^{\!2}\, \1(Z_N\leq1) \bigg] \,\leq\,
\Exv\bigg[ \bigg(\sum_{i=1}^N\log x_i\bigg)^{\!2\,} \bigg] \,\leq\\
&\leq\, N^2\,\Ex\big[\log x\big]^2 \,+\, N\,\Ex\big[(\log x)^2\big] \;.
\end{split} \eeq
In conclusion the lemma is proved splitting $\Exv\big[(\log Z_N)^2\big]$ as $\Exv\big[ (\log Z_N)^2\, \1(Z_N\geq e) \big] +\, \Exv\big[ (\log Z_N)^2\, \1(1<Z_N<e) \big] +\, \Exv\big[ (\log Z_N)^2\, \1(Z_N\leq1) \big]$ and applying the bounds \eqref{eq: bound ZN proof1.3}, \eqref{eq: bound ZN proof2.1}, \eqref{eq: bound ZN proof2.2}.
\endproof

\proof[Proof of the theorem \ref{thm: main}]
It remains to prove only the convergence \eqref{eq: var princ}.
Fix $C<\Phi(\xi^*)\,$.
Fix $0<m<M_C=:M<\infty$ such that \eqref{eq: approx ]-infty,-M] [M,infty[} holds and $m<\xi^*<M\,$: it is possible to make such a choice thanks to the bounds \eqref{eq: bounds xi0} for $\xi^*$ proven in the remark \ref{rk: bounds xi0}.
Fix $\lambda_{m,M}=:\lambda>0$ such that \eqref{eq: approx [-M,-m]} holds.
Let $\varepsilon>0$. Then consider the following random events depending on $x_1,\dots,x_N\,$
\begin{gather*}
E^1_{N,\eps} :=\, \{\,\forall\,\xi\!\in\![0,M]\;\ |\Phi_N(\xi)-\Phi(\xi)|<\varepsilon\,\} \\[4pt]
E^2_{N} :=\, \{\,\forall\,\xi\!\in\![m,M]\;\ \Phi_N(-\xi)<\Phi_N(\xi)-\lambda\,\} \\[4pt]
E^3_{N} :=\, \{\,\forall\,\xi\!\in\![M,\infty[\,\;\ \Phi_N(\xi)<C\,,\,\ \Phi_N(\xi)<\varphi(\xi)\,\}
\end{gather*}
and set $E_{N,\eps}:=E^1_{N,\eps}\cap E^2_{N}\cap E^3_{N}\,$.
It is convenient to split the expectation of $\log Z_N$ as follows:
\eq \label{eq: main proof0}
\Exv\bigg[\frac{1}{N}\log Z_N\bigg] \,=\,
\Exv\bigg[\frac{1}{N}\log Z_N\ \1\big(E_{N,\eps}\big)\bigg] \,+\, \Exv\bigg[\frac{1}{N}\log Z_N\ \1\big((E_{N,\eps})^c\big)\bigg] \;.
\eeq
In the following we are going to see that in the limit $N\to\infty$ the second term on the r.h.s. of \eqref{eq: main proof0} is negligible, while the first term can be computed using the Laplace's method.\\[2pt]
By the lemma \ref{lem: approx}, using the H\"older inequality and the lemma \ref{lem: bound ZN},
\eq \label{eq: main proof1}
\bigg|\, \Exv\bigg[\frac{1}{N}\log Z_N\ \1\big((E_{N,\eps})^c\big)\bigg] \,\bigg| \,\leq\,
\Exv\bigg[\bigg(\frac{1}{N}\log Z_N\bigg)^{\!2\,}\bigg]^{1/2}\; \Probxv\big((E_{N,\eps})^c\big)^{1/2}
\,\xrightarrow[N\to\infty]{}\, 0 \;.
\eeq
[\textit{Upper bound}] Using the Gaussian representation \eqref{eq: gauss repr unif w 1}, a simple upper bound for $Z_N$ is
\eq \label{eq: main proof2.1}
Z_N \,\leq\, \frac{\sqrt{N}}{\sqrt{2\pi w}}\, \int_\R e^{-\frac{N}{2w}\,\xi^2}\, \prod_{i=1}^N|\xi+x_i| \;\dd\xi \,=\,
\frac{\sqrt{N}}{\sqrt{2\pi w}}\, \int_\R e^{N\,\Phi_N(\xi)} \;\dd\xi \;.
\eeq
If the event $E_{N,\eps}$ holds true, remembering also the inequality \eqref{eq: reflect PhiN}, then the following upper bound holds:
\eq \label{eq: main proof2.2} \begin{split}
& \int_\R e^{N\,\Phi_N(\xi)} \;\dd\xi \,\leq\\[2pt]
&\leq\, 2\int_0^m e^{N\,\Phi_N(\xi)} \;\dd\xi \,+\, \int_m^M e^{N\,\Phi_N(\xi)} \;\dd\xi \,+\,
\int_m^M e^{N\,(\Phi_N(\xi)-\lambda)} \;\dd\xi \,+\, 2\int_M^\infty e^{N\,\Phi_N(\xi)} \;\dd\xi \,\leq\\[2pt]
&\leq\, 2\int_0^m e^{N\,(\Phi(\xi)+\eps)} \,\dd\xi \,+\, \int_m^M e^{N\,(\Phi(\xi)+\eps)} \,\dd\xi \,+\,
\int_m^M e^{N\,(\Phi(\xi)+\eps-\lambda)} \,\dd\xi \,+\, 2\,e^{(N-1)\,C}\int_M^\infty e^{\varphi(\xi)} \,\dd\xi \,=\\[2pt]
&\underset{N\to\infty}{=}\, O\big(e^{N\,(\max_{[0,m]}\Phi+\eps)}\big) \,+\, e^{N\,(\Phi(\xi^*)+\eps)}\, \frac{\sqrt{2\pi}\,(1+o(1))}{\sqrt{-N\,\Phi''(\xi^*)}} \,+\, O\big(e^{N\,(\Phi(\xi^*)+\eps-\lambda)}\big) \,+\, O\big(e^{N\,C}\big) \;;
\end{split} \eeq
the last step is obtained by applying the Laplace's method (theorem \ref{thm: laplace}) to the function $\Phi$, which by lemma \ref{lem: Phi} satisfies all the necessary hypothesis.
Now since $\max_{[0,m]}\Phi\,$, $\Phi(\xi^*)-\lambda$ and $C$ are strictly smaller than $\Phi(\xi^*)$, it holds
\eq \label{eq: main proof2.3}
\text{r.h.s. of \eqref{eq: main proof2.2}} \ \,\underset{N\to\infty}{\sim}\,\ e^{N\,(\Phi(\xi^*)+\eps)}\, \frac{\sqrt{2\pi}}{\sqrt{-N\,\Phi''(\xi^*)}} \;.
\eeq
As a consequence of \eqref{eq: main proof2.1}, \eqref{eq: main proof2.2}, \eqref{eq: main proof2.3},
\[ \frac{1}{N}\log Z_N\ \1(E_{N,\eps}) \,\leq\, \Phi(\xi^*)+\eps + O\bigg(\frac{\log N}{N}\bigg) \;,\]
where the $O(\frac{\log N}{N})$ is deterministic. Therefore for all $\eps>0$
\eq \label{eq: main proof2.4}
\limsup_{N\to\infty} \Exv\bigg[\frac{1}{N}\log Z_N\ \1(E_{N,\eps})\bigg] \,\leq\,  \Phi(\xi^*)+\eps \;.
\eeq
[\textit{Lower bound}] Observe that the product $\prod_{i=1}^N(\xi+x_i)$ is always positive for $\xi\geq0$, while it is negative for some $\xi<0$. Hence using the Gaussian representation \eqref{eq: gauss repr unif w 1}, a lower bound for $Z_N$ is
\eq \label{eq: main proof3.1} \begin{split}
Z_N \,&\geq\, \frac{\sqrt{N}}{\sqrt{2\pi w}}\, \bigg( \int_0^\infty e^{-\frac{N}{2w}\,\xi^2}\, \prod_{i=1}^N|\xi+x_i| \;\dd\xi \,- \int_{-\infty}^0 e^{-\frac{N}{2w}\,\xi^2}\, \prod_{i=1}^N|\xi+x_i| \;\dd\xi \bigg) \,=\\[2pt]
&=\, \frac{\sqrt{N}}{\sqrt{2\pi w}}\, \bigg( \int_0^\infty e^{N\,\Phi_N(\xi)} \;\dd\xi \,- \int_{-\infty}^0 e^{N\,\Phi_N(\xi)} \;\dd\xi \bigg) \;.
\end{split} \eeq
If the event $E_{N,\eps}$ holds true, remembering also the inequality \eqref{eq: reflect PhiN}, then the following lower bound holds:
\eq \label{eq: main proof3.2} \begin{split}
& \int_0^\infty e^{N\,\Phi_N(\xi)} \;\dd\xi \,- \int_{-\infty}^0 e^{N\,\Phi_N(\xi)} \;\dd\xi \,\geq\\[2pt]
&\geq\, \int_m^M e^{N\,\Phi_N(\xi)} \;\dd\xi \,- \int_m^M e^{N\,(\Phi_N(\xi)-\lambda)} \;\dd\xi \,\geq\\[2pt]
&\geq\, \int_m^M e^{N\,(\Phi(\xi)-\eps)} \;\dd\xi \,- \int_m^M e^{N\,(\Phi(\xi)+\eps-\lambda)} \;\dd\xi \,=\\
&\underset{N\to\infty}{=}\, e^{N\,(\Phi(\xi^*)-\eps)}\, \frac{\sqrt{2\pi}\,(1+o(1))}{\sqrt{-N\,\Phi''(\xi^*)}} \,-\, e^{N\,(\Phi(\xi^*)+\eps-\lambda)}\, \frac{\sqrt{2\pi}\,(1+o(1))}{\sqrt{-N\,\Phi''(\xi^*)}} \;;
\end{split} \eeq
the last step is obtained by applying the Laplace's method (theorem \ref{thm: laplace}) to the function $\Phi$, which by lemma \ref{lem: Phi} satisfies all the necessary hypothesis.
Now since $\Phi(\xi^*)+\eps-\lambda<\Phi(\xi^*)-\eps$ for all $0<\eps<\frac{1}{2}\lambda\,$, for such a choice of $\eps$ it holds
\eq \label{eq: main proof3.3}
\text{r.h.s. of \eqref{eq: main proof3.2}} \ \,\underset{N\to\infty}{\sim}\,\ e^{N\,(\Phi(\xi^*)-\eps)}\, \frac{\sqrt{2\pi}}{\sqrt{-N\,\Phi''(\xi^*)}} \;.
\eeq
As a consequence of \eqref{eq: main proof3.1}, \eqref{eq: main proof3.2}, \eqref{eq: main proof3.3}, for all $0<\eps<\frac{1}{2}\lambda$
\[ \frac{1}{N}\log Z_N\ \1(E_{N,\eps}) \,\geq\, \bigg(\Phi(\xi^*)-\eps + O\bigg(\frac{\log N}{N}\bigg)\bigg)\; \1(E_{N,\eps}) \;,\]
where the $O(\frac{\log N}{N})$ is deterministic. Therefore, using also the lemma \ref{lem: approx}, for all $0<\eps<\frac{1}{2}\lambda$
\eq \label{eq: main proof3.4}
\liminf_{N\to\infty} \Exv\bigg[\frac{1}{N}\log Z_N\ \1(E_{N,\eps})\bigg] \,\geq\, \liminf_{N\to\infty} \bigg(\Phi(\xi^*)-\eps + O\bigg(\frac{\log N}{N}\bigg)\bigg)\, \Probxv(E_{N,\eps}) \,=\, \Phi(\xi^*)-\eps \;.
\eeq
In conclusion the convergence $\Exv[\frac{1}{N}\log Z_N]\to\Phi(\xi^*)$ as $N\to\infty$ is proven by considering \eqref{eq: main proof0} for $0<\eps<\frac{1}{2}\lambda\,$, then letting $N\to\infty$ exploiting \eqref{eq: main proof1}, \eqref{eq: main proof2.4}, \eqref{eq: main proof3.4}, and finally letting $\eps\to0+$.
\endproof

\begin{rk} \label{rk: determ}
In the deterministic case, namely when the distribution of the $x_i$'s is a Dirac delta centred at a point $x$, the theorem \ref{thm: main} and its corollary \ref{cor: dimer density} reproduce the results obtained in the Proposition 6 of \cite{ACM} by a combinatorial computation.
Indeed the fixed point equation \eqref{eq: fixed point} reduces to $\xi^*=\frac{w}{\xi^*+x}$, whose positive solution is
\[ \xi^* = \frac{-x+\sqrt{x^2+4w}}{2} \;.\]
As a consequence, by \eqref{eq: dimer density} the limiting dimer and monomer density are respectively
\[ d = \frac{(\xi^*)^2}{2w} = \frac{x^2-x\,\sqrt{x^2+4w}+2w}{2w} \;, \quad
m = 1-2\,d = \frac{-x^2+x\,\sqrt{x^2+4w}}{2w} \;.\]
Moreover by \eqref{eq: var princ} and \eqref{eq: dimer density} the limiting pressure can be written as
\[ p = \Phi(\xi^*) = -\frac{(\xi^*)^2}{2w} + \log(\xi^*+x) = - d - \frac{1}{2}\,\log\frac{2\,d}{w} \;.\]
\end{rk}

\section{Self-averaging for monomer-dimer models} \label{sec: self-av}
In this section we prove that under quite general hypothesis a monomer-dimer model with independent random weights has self-averaging pressure density. In particular it will follows that the convergence \eqref{eq: var princ} of the theorem \ref{thm: main} can be strengthen as
\eq \label{eq: var princ as}
\Probxv\text{ - almost surely }\ \exists\,\lim_{N\to\infty} \frac{1}{N}\log Z_N \,=\, \sup_{\xi\geq0}\Phi(\xi) \;,
\eeq
when in the hypothesis of the theorem \ref{thm: main} one substitutes $\Ex[x]<\infty,\,\Ex[(\log x)^2]<\infty$ with the stronger $\Ex[x]<\infty,\,\Ex[x^{-1}]<\infty\,$.

In general let $w_{ij}^{(N)}\geq0\,$, $1\leq i<j\leq N$, $N\in\N$, and $x_i>0\,$, $i\in\N$, be \textit{independent} random variables.
Since the dimer weights may be allowed to take the value $0$ (or to be identically $0$), we do not really know on which kind of graph the model lives, on the contrary the framework is very general (for example the complete graph is included, but also finite-dimensional lattices or diluted random graphs are). This is why we allow a generic dependence of the dimer weights on $N$, in case a normalisation is needed.
During all this section we will denote
\eq
Z_N \,:=\, \sum_{D\in\DD_N}\, \prod_{ij\in D}w_{ij}^{(N)} \!\!\prod_{\,i\in M_N(D)}\!\!\!\!x_i \;.
\eeq
Denote simply by $\E[\,\cdot\,]$ the expectation with respect to all the weights and assume that
\eq \label{eq: self-av hyp}
\sup_N\sup_{1\leq i<j\leq N}\E[w_{ij}^{(N)}]=:C_1<\infty\,, \quad \sup_{i\in\N}\,\E[x_i]=:C_2<\infty\,, \quad \sup_{i\in\N}\,\E[x_i^{-1}]=:C_3<\infty \;.
\eeq
Clearly the pressure $p_N:=\frac{1}{N}\log Z_N$ is a random variable and it has finite expectation, indeed
\[ N\,p_N \,\begin{cases}
\,\geq\, \log \prod_{i=1}^N x_i \,=\, \sum_{i=1}^N \log x_i \,\geq\, \sum_{i=1}^N (1+x_i^{-1}) \; \in L^1(\Prob)\\
\,\leq\, Z_N-1\; \in L^1(\Prob) \end{cases} \,. \]
The following theorem shows that in the limit $N\to\infty$ the pressure $p_N$ concentrates around its expectation, or in other terms it tends to become a deterministic quantity.

\begin{thm} \label{thm: self-av}
Let $w_{ij}^{(N)}\geq0\,$, $1\leq i<j\leq N$, $N\in\N$, and $x_i>0\,$, $i\in\N$, be \textit{independent} random variables
that satisfy \eqref{eq: self-av hyp}.
Then for all $t>0$, $N\in\N$, $q\geq1$
\begin{equation} \label{eq: self-av1}
\Prob\big( \,|p_N-\E[p_N]| \geq t \big) \,\leq\,
2\,\exp\bigg(-\frac{t^2\,N}{4\,q^2\,\log^2 N} \bigg) \,+\, (a+b\,N)\,N^{1-q} \;,
\end{equation}
where $a:=4+2C_2C_3\,$, $b:=2C_1C_3^2\,$.
As a consequence, choosing $q>3$,
\begin{equation} \label{eq: self-av2}
|p_N-\E[p_N]| \xrightarrow[N\to\infty]{} 0\,\ \Prob\text{-almost surely} \;.
\end{equation}
\end{thm}

If the random variables $w_{ij}^{(N)},\,x_i,\,x_i^{-1}$ are bounded, then one could obtain an exponential rate of convergence instead of \eqref{eq: self-av1}, but here we prefer to obtain the result \eqref{eq: self-av2} with minimal assumptions.

\proof
Fix $N\in\N$. Set $w_i:=\big(w_{i(i+1)}^{(N)},\dots,w_{iN}^{(N)}\big)$ for all $i=1,\dots,N-1\,$.
We consider the filtration of length $2N-1$ such that in the first $N$ steps the monomer weights $x_i$ are exposed, while in the last $N-1$ steps the vectors $w_i$ of dimer weights are exposed.
Since $p_N$ is a function of $x_1,\dots,x_N$, $w_1,\dots,w_{N-1}$ and $\E[|p_N|]<\infty$, we may define the Doob martingale of $p_N$ with respect to this filtration
\begin{gather*}
M_i := \E\big[\,p_N\,\big|\,x_1,\dots,x_i\big] \quad\forall\,i=0,\dots,N \;,\\[4pt]
M_{N+i} := \E\big[\,p_N\,\big|\,x_1,\dots,x_N,w_1,\dots,w_i\big] \quad\forall\,i=1,\dots,N-1 \;;
\end{gather*}
in particular it holds $M_0=\E[\,p_N]$ and $M_{2N-1}=p_N$.\\
Now we want to bound the increments $|M_i-M_{i-1}|$ for every $i=1,\dots,2N-1$, in order to apply the Azuma's inequality.
By hypothesis $x_1,\dots,x_N$, $w_1,\dots,w_{N-1}$ are stochastically independent, hence the conditional expectations are simply $M_i=\E_{\x^{i+1},\,\w}[p_N]$ for $i=0,\dots,N$ and $M_{N+i}=\E_{\w^{i+1}}[p_N]$ for $i=1,\dots,N-1\,$. 
As a consequence it is easy to check that for $i=1,\dots,N$ it holds
\begin{equation} \label{eq: self-av proof1.1}
|M_i-M_{i-1}| \,\leq\, \sup_{\tilde\x_{i-1},\,\tilde\x^{i+1},\,\tilde\w} \big|\, p_N\big(\tilde\x_{i-1},\,x_i,\,\tilde\x^{i+1},\,\tilde\w\big) - \E_{x_i}\big[\, p_N\big(\tilde\x_{i-1},\,x_i,\,\tilde\x^{i+1},\,\tilde\w\big) \big] \big| \end{equation}
and for $i=1,\dots,N-1$ it holds
\begin{equation} \label{eq: self-av proof1.2}
|M_{N+i}-M_{N+i-1}| \,\leq\, \sup_{\tilde\w_{i-1},\,\tilde\w^{i+1}} \big|\, p_N\big(\x,\,\tilde\w_{i-1},\,w_i,\,\tilde\w^{i+1}\big) - \E_{w_i}\big[\, p_N\big(\x,\,\tilde\w_{i-1},\,w_i,\,\tilde\w^{i+1}\big) \big] \big| \;.
\end{equation}
Here we have adopted the following notation $\x:=(x_1,\dots,x_N)$, $\x_k:=(x_1,\dots,x_k)$, $\x^k:=(x_k,\dots,x_N)$ and similarly $\w:=(w_1,\dots,w_{N-1})$, $\w_k:=(w_1,\dots,w_k)$, $\w^k:=(w_k,\dots,w_N)$; the symbols with a tilde denote a deterministic value taken by the corresponding random quantity.\\

First fix $i=1,\dots,N$, fix the deterministic vectors $\tilde\x_{i-1},\,\tilde\x^{i+1},\,\tilde\w$ and let $x_i',\,x_i''$ be two independent random variables distributed as $x_i$. Set
\[ p_N':=p_N\big(\tilde\x_{i-1},x_i',\,\tilde\x^{i+1},\,\tilde\w\big)\,,\quad
p_N'':=p_N\big(\tilde\x_{i-1},\,x_i'',\,\tilde\x^{i+1},\,\tilde\w\big) \;.\]
To estimate the difference between $p_N',\,p_N''$ we use the Heilmann-Lieb recursion for the partition function of a monomer-dimer model (see \cite{HL} and the proposition \ref{prop: HL rec}):
\begin{equation} \label{eq: self-av proof2.1} \begin{split}
p_N'-p_N'' \,&=\, \frac{1}{N} \log \frac{Z_N'}{Z_N''} \,=\,
\frac{1}{N} \log \frac{ x_i'\, Z_{-i} \,+\, \sum_{j=1}^{i-1} \tilde w_{ji}\, Z_{-j-i} \,+\, \sum_{j=i+1}^N \tilde w_{ij}\, Z_{-i-j} } { x_i''\, Z_{-i} \,+\, \sum_{j=1}^{i-1} \tilde w_{ji}\, Z_{-j-i} \,+\, \sum_{j=i+1}^N \tilde w_{ij}\, Z_{-i-j}} \,\leq\\[2pt]
&\leq\, \frac{1}{N} \log\bigg( \frac{x_i'}{x_i''} \,+\, 1 \bigg) \;;
\end{split} \end{equation}
here we denote by $Z_{-i},\,Z_{-i-j}$ the partitions function of the model over the vertices $\{1,\dots,N\}\smallsetminus\{i\}$, $\{1,\dots,N\}\smallsetminus\{i,j\}$ respectively, with weights $\tilde\x_{i-1},\,\tilde\x^{i+1},\,\tilde\w_{i-1},\,\tilde\w^{i+1}$. It is important (for the inequality in \eqref{eq: self-av proof2.1}) to notice that these partition functions do not depend on the weights $x_i',\,x_i''$.
In the same way one finds
\begin{equation} \label{eq: self-av proof2.2} \begin{split}
p_N''-p_N' \,\leq\, \frac{1}{N} \log\bigg( \frac{x_i''}{x_i'} \,+\, 1 \bigg) \;.
\end{split} \end{equation}
Denote by $\E''$ the expectation with respect to the random variable $x_i''$ only. Then the inequalities \eqref{eq: self-av proof2.1}, \eqref{eq: self-av proof2.2} provide respectively the following random bounds
\begin{gather} \label{eq: self-av proof2.3}
p_N'-\E[p_N''] \,=\, \E''[p_N'-p_N''] \,\overset{\eqref{eq: self-av proof2.1}}{\leq}\,
\E''\bigg[ \frac{1}{N} \log\bigg( \frac{x_i'}{x_i''} \,+\, 1 \bigg) \bigg] \,\leq\,
\frac{1}{N} \log\big( x_i'\,\E[x_i^{-1}] + 1 \big) \;;\\[2pt]
\label{eq: self-av proof2.4}
\E[p_N'']-p_N' \,=\, \E''[p_N''-p_N'] \,\overset{\eqref{eq: self-av proof2.2}}\leq\,
\E''\bigg[ \frac{1}{N} \log\bigg( \frac{x_i''}{x_i'} \,+\, 1 \bigg) \bigg] \,\leq\,
\frac{1}{N} \log\big( \E[x_i]\,(x_i')^{-1} + 1 \big) \;.
\end{gather}
Choose $q>0$ and the previous inequalities provide a bound for $|M_i-M_{i-1}|$ that holds true ``with high probability'':
\begin{equation} \label{eq: self-av proof2.5} \begin{split}
&\Prob\bigg( |M_i-M_{i-1}| > \frac{q}{N}\log N \bigg) \,\overset{\eqref{eq: self-av proof1.1}}{\leq}\,
\Prob\bigg( \sup_{\,\tilde\x_{i-1},\,\tilde\x^{i+1},\,\tilde\w}\!\big|p_N'-\E[p_N'']\big| > \frac{q}{N}\log N \bigg) \,\leq\\[2pt]
&\leq\, \Prob\bigg( \sup_{\,\tilde\x_{i-1},\,\tilde\x^{i+1},\,\tilde\w}\!\!\big(p_N'-\E[p_N'']\big) > \frac{q}{N}\log N \bigg) \,+\,
\Prob\bigg( \sup_{\,\tilde\x_{i-1},\,\tilde\x^{i+1},\,\tilde\w}\!\!\big(\E[p_N'']-p_N'\big) > \frac{q}{N}\log N \bigg) \,\overset{\eqref{eq: self-av proof2.3},\eqref{eq: self-av proof2.4}}{\leq}\\[2pt]
&\leq\, \Prob\bigg( \frac{1}{N} \log\big( x_i\,\E[x_i^{-1}] + 1 \big) > \frac{q}{N}\log N \bigg) \,+\,
\Prob\bigg( \frac{1}{N} \log\big( \E[x_i]\,x_i^{-1} + 1 \big) > \frac{q}{N}\log N \bigg) \,=\\[2pt]
&=\, \Prob\bigg( 1+x_i\,\E[x_i^{-1}] > N^q \bigg) \,+\,
\Prob\bigg( 1+\E[x_i]\,x_i^{-1} > N^q \bigg) \,\leq\\[2pt]
&\leq\, \E\bigg[ 1+x_i\,\E[x_i^{-1}] \bigg] \, N^{-q} \,+\,
\E\bigg[ 1+\E[x_i]\,x_i^{-1} \bigg] \, N^{-q} \,\leq\\[2pt]
&\leq\, 2\,(1+C_2C_3)\,N^{-q} \;;
\end{split} \raisetag{4\baselineskip} \end{equation}
here at the penultimate step we have used the Markov inequality.\\

Now instead fix  $i=1,\dots,N-1$, fix the deterministic vectors $\tilde\w_{i-1},\,\tilde\w^{i+1}$, let $w_i',\,w_i''$ be two independent random vectors distributed as $w_i$ and leave the vector of monomer weights $\x$ random (choose $w_i',\,w_i''$ independent of $\x$ too).
Reassign the notation previously used, setting now:
\[ p_N':=p_N\big(\x,\,\tilde\w_{i-1},\,w_i',\,\tilde\w^{i+1}\big)\,,\quad
p_N'':=p_N\big(\x,\,\tilde\w_{i-1},\,w_i'',\,\tilde\w^{i+1}\big) \;.\]
To estimate the difference between $p_N',\,p_N''$ we use again the Heilmann-Lieb recursion for the partition function (see \cite{HL} and the proposition \ref{prop: HL rec}):
\begin{equation} \label{eq: self-av proof3.1} \begin{split}
p_N'-p_N'' \,&=\, \frac{1}{N} \log \frac{Z_N'}{Z_N''} \,=\,
\frac{1}{N} \log \frac{ x_i\, Z_{-i} \,+\, \sum_{j=1}^{i-1} \tilde w_{ji}\, Z_{-j-i} \,+\, \sum_{j=i+1}^N w'_{ij}\, Z_{-i-j} } { x_i\, Z_{-i} \,+\, \sum_{j=1}^{i-1} \tilde w_{ji}\, Z_{-j-i} \,+\, \sum_{j=i+1}^N w''_{ij}\, Z_{-i-j}} \,\leq\\[2pt]
&\leq\, \frac{1}{N} \log\bigg( 1 \,+\, \frac{ \sum_{j=i+1}^N w'_{ij}\, Z_{-i-j} } {x_i\, Z_{-i} } \bigg) \,=\,
\frac{1}{N} \log\bigg( 1 \,+ \sum_{j=i+1}^N \frac{w'_{ij}}{x_i\,x_j}\, \langle\1_{j\in M}\rangle_{-i} \bigg) \,\leq\\
&\leq\, \frac{1}{N} \log\bigg( 1 \,+ \sum_{j=i+1}^N \frac{w'_{ij}}{x_i\,x_j} \bigg) \;;
\end{split} \end{equation}
we have denoted by $Z_{-i},\,Z_{-i-j}$ the partitions function of the model over the vertices $\{1,\dots,N\}\smallsetminus\{i\}$, $\{1,\dots,N\}\smallsetminus\{i,j\}$ respectively, with weights $\x_{i-1},\,\x^{i+1},\,\tilde\w_{i-1},\,\tilde\w^{i+1}$. It is important (for the first inequality in \eqref{eq: self-av proof3.1}) to notice that these partition functions do not depend on the weights $w_i',\,w_i''$.
In the same way one finds
\begin{equation} \label{eq: self-av proof3.2} \begin{split}
p_N''-p_N' \,\leq\, \frac{1}{N} \log\bigg( 1 \,+ \sum_{j=i+1}^N \frac{w''_{ij}}{x_i\,x_j} \bigg) \;.
\end{split} \end{equation}
Denote by $\E''$ the expectation with respect to the random vector $w_i''$ only. Then the inequalities \eqref{eq: self-av proof3.1},  \eqref{eq: self-av proof3.2} provide respectively the following random bounds
\begin{gather} \label{eq: self-av proof3.3}
p_N'-\E''[p_N''] \,=\, \E''[p_N'-p_N''] \,\overset{\eqref{eq: self-av proof3.2}}{\leq}\,
\frac{1}{N} \log\bigg( 1 \,+ \sum_{j=i+1}^N \frac{w'_{ij}}{x_i\,x_j} \bigg) \;;\\[2pt]
\label{eq: self-av proof3.4} \begin{split}
\E''[p_N'']-p_N' \,&=\, \E''[p_N''-p_N'] \,\overset{\eqref{eq: self-av proof3.3}}\leq\,
\E''\bigg[\frac{1}{N} \log\bigg( 1 \,+ \sum_{j=i+1}^N \frac{w''_{ij}}{x_i\,x_j} \bigg)\bigg] \,\leq\\[-2pt]
&\leq\, \frac{1}{N} \log \bigg( 1 \,+ \sum_{j=i+1}^N \frac{\E[w_{ij}]}{x_i\,x_j} \bigg) \;. \end{split}
\end{gather}
Choose $q>0$ and the previous inequalities provide a bound for $|M_{N+i}-M_{N+i-1}|$ that holds true ``with high probability'':
\begin{equation} \label{eq: self-av proof3.5} \begin{split}
&\Prob\bigg( |M_{N+i}-M_{N+i-1}| > \frac{q}{N}\log N \bigg) \,\overset{\eqref{eq: self-av proof1.2}}{\leq}\,
\Prob\bigg( \sup_{\tilde\w_{i-1},\,\tilde\w^{i+1}}\!\big|p_N'-\E''[p_N'']\big| > \frac{q}{N}\log N \bigg) \,\leq\\
&\leq\, \Prob\bigg( \sup_{\tilde\w_{i-1},\,\tilde\w^{i+1}}\!\!\big(p_N'-\E''[p_N'']\big) > \frac{q}{N}\log N \bigg) \,+\,
\Prob\bigg( \sup_{\tilde\w_{i-1},\,\tilde\w^{i+1}}\!\!\big(\E''[p_N'']-p_N'\big) > \frac{q}{N}\log N \bigg) \,\overset{\eqref{eq: self-av proof3.2},\eqref{eq: self-av proof3.3}}{\leq}\\
&\leq\, \Prob\bigg( \frac{1}{N} \log\big(1+\!\sum_{j=i+1}^N\frac{w_{ij}}{x_i\,x_j}\big) > \frac{q}{N}\log N \bigg) \,+\,
\Prob\bigg( \frac{1}{N} \log \big(1+\!\sum_{j=i+1}^N\frac{\E[w_{ij}]}{x_i\,x_j}\big) > \frac{q}{N}\log N\,\bigg) \,=\\
&\leq\, \Prob\bigg( 1+\!\sum_{j=i+1}^N\frac{w_{ij}}{x_i\,x_j} > N^q \bigg) \,+\,
\Prob\bigg( 1+\!\sum_{j=i+1}^N\frac{\E[w_{ij}]}{x_i\,x_j} > N^q \bigg) \,\leq\\
&\leq\, \E\bigg[\, 1+\!\!\sum_{j=i+1}^N\frac{w_{ij}}{x_i\,x_j} \,\bigg] \, N^{-q} \,+\,
\E\bigg[\, 1+\!\!\sum_{j=i+1}^N\frac{\E[w_{ij}]}{x_i\,x_j} \,\bigg] \, N^{-q} \,\leq\\[4pt]
&\leq\, 2\, (1+N\,C_1C_3^2)\, N^{-q} \;;
\end{split} \raisetag{4\baselineskip} \end{equation}
here at the penultimate step we have applied the Markov inequality.\\

As an immediate consequence of \eqref{eq: self-av proof2.5} and \eqref{eq: self-av proof3.5},
\begin{equation} \label{eq: self-av proof4.1} \begin{split}
& \Prob\bigg( \exists\,i=1,\dots,{2N-1}\text{ s.t. }|M_i-M_{i-1}| > \frac{q}{N}\log N \bigg) \,\leq\\
&\leq\, N\,\big(2\,(1+C_2C_3)\,N^{-q}\big) \,+\, (N-1)\,\big(2\,(1+N\,C_1C_3^2)\,N^{-q}\big) \\[2pt]
&\leq\, 2\,\big(2+C_2C_3+C_1C_3^2\,N\big)\,N^{1-q} \;.
\end{split} \end{equation}
Therefore by the extended Azuma's inequality (theorem \ref{thm: azuma extended}), for all $t>0$ it holds
\begin{equation} \label{eq: self-av proof4.2}
\Prob\big( |M_{N-1}-M_0| \geq t \big) \,\leq\, 2\, \exp\bigg(\! -\frac{t^2}{2}\,\frac{N}{2\,q^2\,\log^2 N} \bigg) \,+\, 2\,\big(2+C_2C_3+C_1C_3^2\,N\big)\,N^{1-q}
\end{equation}
and the proof of \eqref{eq: self-av1} is concluded.
Choosing $q>3$ the r.h.s. of \eqref{eq: self-av1} is summable with respect to $N\in\N$, hence \eqref{eq: self-av2} follows by a standard application of the Borel-Cantelli lemma.
\endproof

\section*{Appendix}
\renewcommand{\thethm}{A\arabic{thm}}
\setcounter{thm}{0}
\renewcommand{\theequation}{A\arabic{equation}}
\setcounter{equation}{0}

In this appendix we state the main technical results used in the paper. We omit their proofs, that can be found in the literature.

\begin{thm}[Gaussian integration by parts; Wick-Isserlis formula] \label{thm: gauss calculus}
Let $(\xi_1,\dots,\xi_n)$ be a Gaussian random vector with mean $0$ and positive semi-definite covariance matrix $C=(c_{ij})_{i,j=1,\dots,n}\,$.
Let $f\!:\R^{n-1}\to\R$ be a differentiable function such that $\E\big[\big|\xi_1\,f(\xi_2,\dots,\xi_n)\big|\big]<\infty$ and $\E\big[\big|\frac{\partial f}{\partial \xi_j}(\xi_2,\dots,\xi_n)\big|\big]<\infty$ for all $j=2,\dots,n$.
Then:
\eq \label{eq: integr by parts}
\E\big[\xi_1\,f(\xi_2,\dots,\xi_n)\big] \,=\, \sum_{j=2}^n\, c_{1j}\, \E\bigg[\frac{\partial f}{\partial \xi_j}(\xi_2,\dots,\xi_n)\bigg] \;.
\eeq
As a consequence one can prove the following:
\eq \label{eq: wick}
\E\bigg[\prod_{i=1}^n\xi_i\bigg] \;= \!\sum_{P\text{ partition of}\atop\{1,\dots,n\}\text{ into pairs}}\, \prod_{\{i,j\}\in P} c_{ij} \;.
\eeq
\end{thm}

The Gaussian integration by parts \eqref{eq: integr by parts} can be found in \cite{T}.
The Wick-Isserlis formula \eqref{eq: wick} follows by \eqref{eq: integr by parts} using an induction argument; but it appeared for the first time in \cite{I}.

\begin{thm}[Laplace's method] \label{thm: laplace}
Let $\phi\!:[a,b]\to\R$ be a function of class $C^2$. Suppose that there exists $x_0\in\,]a,b[\,$ such that
\begin{itemize}
\item[i.] $\phi(x_0)>\phi(x)$ for all $x\in[a,b]\,$ (i.e. $x_0$ is the only global maximum point of $\phi$);
\item[ii.] $\phi''(x_0)<0\,$.
\end{itemize}
Then as $n\to\infty$
\eq \label{eq: laplace}
\int_a^b e^{n\,\phi(x)}\, \dd x \,=\,
e^{n\,\phi(x_0)}\; \frac{\sqrt{2\pi}}{\sqrt{-n\,\phi''(x_0)}}\; \big(1+o(1)\big) \;.
\eeq
\end{thm}

A formal proof of the Laplace's method can be found in \cite{Db}.

\begin{thm}[uniform weak law of large numbers] \label{thm: uwlln}
Let $\bigchi,\,\Theta$ be metric spaces.
Let $X_i,\,i\in\N$ be i.i.d.$\!$ random variables taking values in $\bigchi$.
Let $f\!:\bigchi\times\Theta\to\R$ be a function such that $f(\cdot,\theta)$ is measurable for all $\theta\in\Theta$.
Suppose that:
\begin{itemize}
\item[i.] $\Theta$ is compact;
\item[ii.] $\Prob\big(f(X_1,\cdot)\!\text{ is$\!$ continuous$\!$ at }\theta\big)=1\ $ for all $\theta\in\Theta\,$;
\item[iii.] $\exists$ $F\!:\bigchi\to[0,\infty]$ such that $\Prob\big(|f(X_1,\theta)|\leq F(X_1)\big)=1$ for all $\theta\in\Theta$ and $\E[F(X_1)]<\infty\,$.
\end{itemize}
Then for all $\eps>0$
\eq \label{eq: uwlln}
\Prob\bigg(\, \sup_{\theta\in\Theta}\, \bigg|\,\frac{1}{n}\sum_{i=1}^n f(X_i,\theta)-\E[f(X,\theta)]\,\bigg| \,\geq \varepsilon \,\bigg) \,\xrightarrow[n\to\infty]{}\, 0  \;.
\eeq
\end{thm}

The uniform law of large number appeared in \cite{J}. It is based on the (standard) law of large numbers and on a compactness argument.

\begin{thm}[extension of the Azuma's inequality] \label{thm: azuma extended}
Let $M=(M_i)_{i=0,\dots,n}$ be a real martingale with respect to a filter.
Suppose that there exist constants $\eps>0$ and $c_1,\dots,c_n<\infty$ such that
\[ \Prob\big(\exists\,i=1,\dots,n\textup{ s.t. }|M_i-M_{i-1}|>c_i\big) \,\leq\, \eps \;.\]
Then for all $t>0$
\eq \label{eq: azuma extended}
\Prob\big(|M_n-M_0|>t\big) \,\leq\, 2\,\exp\bigg(\!-\frac{t^2}{2\,\sum_{i=1}^n c_i^2}\bigg) +\, \eps \;.
\eeq
\end{thm}

The Azuma's inequality is a useful tool in the martingale theory that allows to obtain concentration results. Its usual formulation is given with $\eps=0$. The extension with $\eps>0$ can be found in \cite{CL}; but it can be proven also starting from the usual formulation and introducing a suitable stopping time, following the ideas in \cite{W}.



\end{document}